\documentclass[conference]{IEEEtran}
\IEEEoverridecommandlockouts
\usepackage{cite}
\usepackage{amsmath,amssymb,amsfonts}
\usepackage{algorithmic}
\usepackage{graphicx}
\usepackage{textcomp}
\usepackage{xcolor}
\usepackage{tabularx}
\usepackage{booktabs}
\usepackage{setspace}
\usepackage{multirow}
\def\BibTeX{{\rm B\kern-.05em{\sc i\kern-.025em b}\kern-.08em
    T\kern-.1667em\lower.7ex\hbox{E}\kern-.125emX}}
\begin{document}

\title{Bayesian-Driven Graph Reasoning for Active Radio Map Construction\\
{}
}
\author{
\IEEEauthorblockN{
Wenlihan Lu\IEEEauthorrefmark{1},
Shijian Gao\IEEEauthorrefmark{1},
Miaowen Wen\IEEEauthorrefmark{2},
Yuxuan Liang\IEEEauthorrefmark{1},
Liuqing Yang\IEEEauthorrefmark{1},
Chan-Byoung Chae\IEEEauthorrefmark{3},
H. Vincent Poor\IEEEauthorrefmark{4}
}
\IEEEauthorblockA{
\IEEEauthorrefmark{1} Hong Kong University of Science and Technology (Guangzhou), Guangzhou, China \\
\IEEEauthorrefmark{2} South China University of Technology, Guangzhou, China \\
\IEEEauthorrefmark{3} Yonsei University, Seoul, Korea\\
\IEEEauthorrefmark{4} Princeton University, NJ, USA\\
wlu162@connect.hkust-gz.edu.cn, \{shijiangao, yuxuanliang, lqyang\}@hkust-gz.edu.cn, eemwwen@scut.edu.cn, \\
cbchae@yonsei.ac.kr, poor@princeton.edu
}
}

\maketitle
\newcommand{\dist}[2]{\left\lVert #1-#2 \right\rVert_2} 
\begin{abstract}
\setstretch{0.8}
With the emergence of the low-altitude economy, radio maps have become essential for ensuring reliable wireless connectivity to aerial platforms. Autonomous aerial agents are commonly deployed for data collection using waypoint-based navigation; however, their limited battery capacity significantly constrains coverage and efficiency. To address this, we propose an uncertainty-aware radio map (URAM) reconstruction framework that explicitly leverages graph-based reasoning tailored for waypoint navigation. Our approach integrates two key deep learning components: (1) a Bayesian neural network that estimates spatial uncertainty in real time, and (2) an attention-based reinforcement learning policy that performs global reasoning over a probabilistic roadmap, using uncertainty estimates to plan informative and energy-efficient trajectories. This graph-based reasoning enables intelligent, non-myopic trajectory planning, guiding agents toward the most informative regions while satisfying safety constraints. Experimental results show that URAM improves reconstruction accuracy by up to 34\% over existing baselines.\\

\end{abstract}


\vspace{-2mm}
\setstretch{0.82}
\section{Introduction}

The rise of the low-altitude economy has intensified the need for reliable wireless connectivity in near-ground airspace. To support this, radio map, which can also be understood more broadly as a type of Channel Knowledge Map (CKM) \cite{ckm}, play a key role by providing spatial representations of channel characteristics such as received signal strength (RSS), channel gain, and power spectral density, enabling effective network planning and aerial operation \cite{radiomap_survey}.

Despite being a widely studied tool for network optimization, obtaining high-quality radio maps remains challenging in practice, particularly in the era of the low-altitude economy, where the spatial dimension has become more expansive than ever. Existing approaches face significant practical limitations. Ray-tracing methods \cite{raytracing} offer physical accuracy but are computationally expensive and require detailed environmental models, limiting their scalability. Data-driven methods such as kriging \cite{kriging}, matrix completion \cite{matrixc} and deep learning \cite{autoencoder, nerf} support efficient reconstruction but inherently depend on pre-collected measurements. The acquisition of such data is often resource-prohibitive. Manual surveys are characterized by significant labor requirements, whereas autonomous aerial agents are fundamentally constrained by limited battery endurance and payload capacity \cite{payload}. These constraints are further exacerbated by the weight and power demands of onboard instrumentation (e.g., spectrum analyzers). Consequently, the development of active, online data acquisition strategies is necessitated, particularly for low-altitude networks where accurate radio maps significantly enhance navigation reliability, interference management, and coverage optimization for aerial agents. The principal challenge lies in the intelligent formulation of an agent's trajectory to maximize the collection of information-rich data under finite resource constraints.

In light of the aforementioned issues, a growing body of research has explored active radio map construction, aiming to select sampling locations that maximise information gain; however, existing approaches exhibit key limitations in uncertainty modeling and planning adaptability.
Methods based on Gaussian Processes (GPs) \cite{activeGP, ipp} jointly model signal distribution and  predictive variance to guide sampling, but they suffer from poor reconstruction performance.
Alternatively, while many works employ a separate neural network trained specifically to predict the discrepancy between generated outputs and ground truth \cite{spectrumsurveying, oreman}, these approaches often lack generalization.
Furthermore, common planning strategies like greedy next-place selection based on expected information gain~\cite{spectrumsurveying, oreman} often result in myopic decisions and poor adaptability.
Critically, prior works frequently overlook or only implicitly address real-world constraints such as limited energy, environmental obstacles, and fixed start-to-goal trajectories—factors that significantly hinder practical deployment. 

\begin{figure*}[!t]
    \centering
    \includegraphics[width=1.0\linewidth]{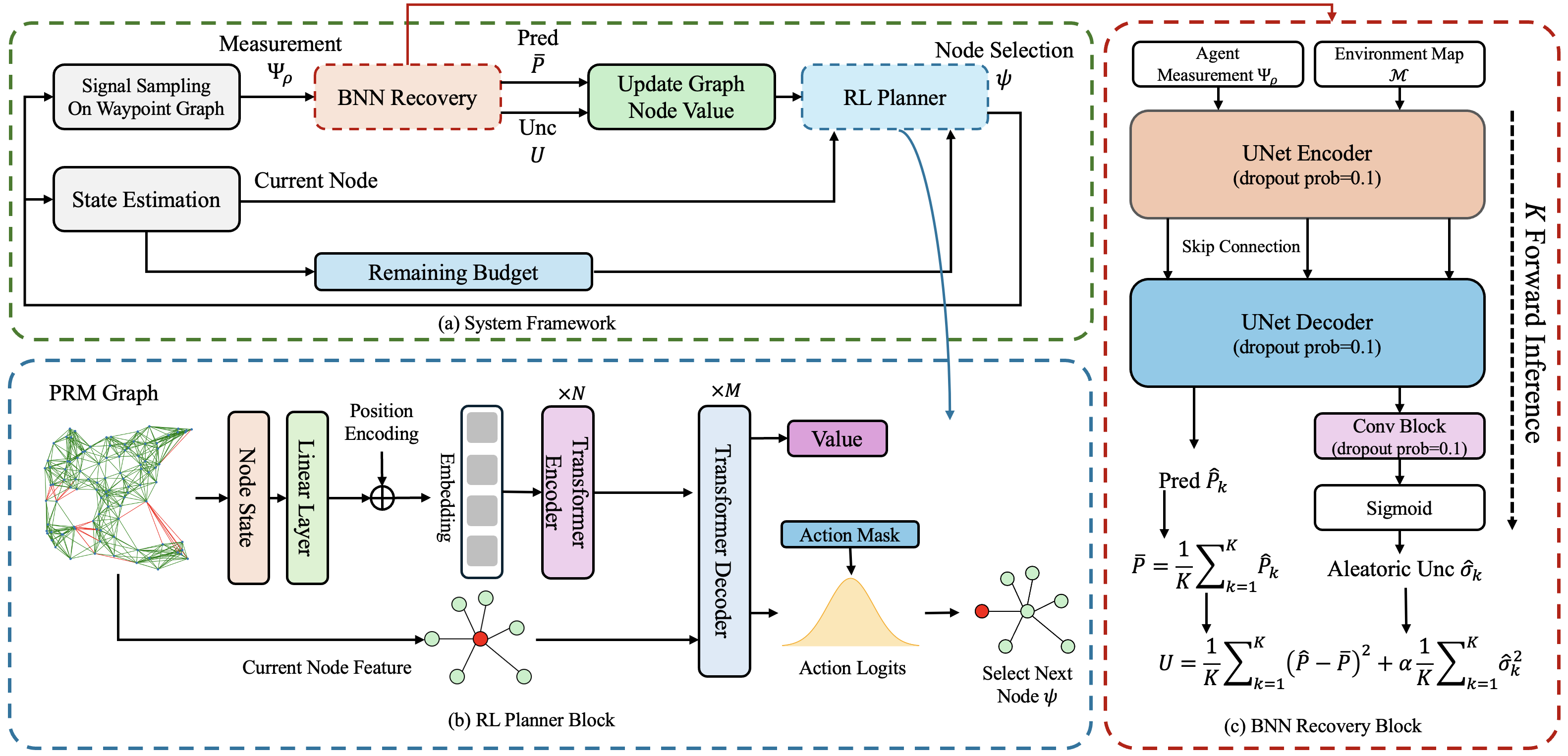}
    \caption{An illustration of the proposed uncertainty-aware radio map construction framework.}
    \label{system}
\end{figure*}

To address these limitations, we propose an uncertainty-aware radio map (URAM) construction framework. URAM aims to integrate deep uncertainty modeling with RL-based path planning to navigate between specified start and goal points under a strict mission budget. Two key design choices ground our framework in practical efficacy.
First, to align with real-world applications \cite{cao2022catnipp}, we transition from continuous trajectory planning to graph-based reasoning. Recognizing that navigation systems of most aerial agents are waypoint-based, we construct Probabilistic Roadmap (PRM) \cite{prm} to serve as the foundation for our budget-aware, goal-oriented planner.
Second, for uncertainty modelling, we leverage a Bayesian U-Net. This architectural choice is built on \cite{sip2net}, which established its outstanding performance in accurately reconstructing radio maps. By incorporating Bayesian principles through heteroscedastic regression and Monte Carlo (MC) dropout, our model reliably quantifies both epistemic and aleatoric uncertainty to guide exploration.
Together, these components form an iterative, closed-loop framework for efficient, uncertainty-aware online radio map reconstruction.

In summary, this paper presents the URAM framework for active radio map construction, featuring a closed-loop system that combines a Bayesian Neural Network (BNN) with a  graph reasoning agent. This design enables efficient data acquisition under strict operational constraints. Experiments demonstrate that URAM can achieve over a 30\% improvement in terms of reconstruction accuracy over existing baselines.

\setstretch{0.8}
\section{Problem Statement}
This section formalizes the active radio map reconstruction task. Our goal is to select a trajectory that enables accurate signal reconstruction with minimal resource usage.
Without loss of generality, we consider a 2D spatial radio map $P$ that captures signal distribution in a single frequency band under quasi-static conditions. This simplified setting allows for a detailed exploration of uncertainty estimation and decision-making under travel and navigation constraints. 

\subsection{Sampling Model}
Consider an autonomous agent that explores an environment with a known geographic map~$\mathcal{M}$, a starting point~$s$, and a goal point~$g$, with a given travel budget~$B$ (e.g., battery capacity). Let
$\mathcal{M}\!\in\!\{0,1\}^{N\times N}$ be a binary map of the operating area, in which
$\mathcal{M}(x,y)=0$ denotes free space and $\mathcal{M}(x,y)=1$ denotes an obstacle.
The unknown ground-truth power field is
$P\in\mathbb{R}^{N \times N}$.

The agent traverses an ordered sequence of waypoints
$
\rho \triangleq \bigl\langle v_1,v_2,\dots,v_n \bigr\rangle
$, with fixed starting point and ending point, i.e., $
v_1=s$ and $v_n=g$.
Let $\mathcal{F} = \{(x,y)\mid \mathcal{M}(x,y)=0\}$ denotes the set of obstacle-free positions, then the $i$-th point within $\rho$, denoted as  $\rho_i$ must belong to  $\mathcal{F}$. The agent moves along straight-line segments connecting consecutive waypoints, ensuring that each segment remaining entirely within $\mathcal{F}$.

Let $\text{seg}(x, y)$ represent the segment connecting point $x$ to point $y$. While traversing $\text{seg}(v_i, v_{i+1})$, the agent acquires a dense set of signal measurements along the path. The full set of collected observations is represented as
\[
\Psi_\rho \triangleq
\bigl\{ \bigl((x,y),P(x,y)\bigr)\mid (x,y)\in\text{seg}(v_i,v_{i+1}) \bigr\}.
\]

\subsection{Problem Formulation}
Given the sampled measurements $\Psi_\rho$ and a known environmental map $\mathcal{M}$, a neural estimator $f_\theta$ is used to predict the complete radio map
\[
\hat{P} = f_\theta\bigl(\Psi_\rho, \mathcal{M}\bigr).
\]
Let $\mathcal{E}(\cdot,\cdot)$ denote a task-level error metric, such as mean squared error (MSE). The objective is to select a trajectory $\rho$ that minimizes the reconstruction error by solving the following problem:
\[
\begin{aligned}
\min_{\rho}\quad &
\mathcal{E}\Bigl(P,\hat{P}\bigl(\Psi_\rho\bigr)\Bigr) \\[2pt]
\text{s.t.}\quad &
C(\rho)\le B,\quad \\
&\rho_1 = s,\\&\rho_n = g .
\end{aligned}
\]
Here, the trajectory cost $C(\rho)$ is defined as the cumulative segment cost: $
C(\rho)\propto \sum_{i=1}^{n-1} \dist{v_i}{v_{i+1}},
$
where $\dist{v_i}{v_{i+1}}$ is the Euclidean distance between consecutive waypoints, which is a common choice for measuring cost without loss of generality.

\section{Proposed Framework}
To support efficient, uncertainty-aware sensing, URAM integrates reconstruction and planning in a closed-loop system, as illustrated in Fig. \ref{system}. Given the agent’s current measurements and the environment map, a Bayesian U-Net estimates the global radio map and its uncertainty. These predictions update the waypoint graph node values, which are then used by a transformer-based RL planner to select the next sampling node. This process iterates within a budget constraint, enabling adaptive and uncertainty-aware exploration.

\subsection{Radio Map Prediction with Uncertainty Output}

To reconstruct the full radio map from partial observations, we adopt a U-Net-based deep neural network architecture. U-Net has proven to be highly effective in radio map construction due to its ability to capture both fine-grained local patterns and broad spatial context through its encoder--decoder structure with skip connections. To improve computational efficiency, we employ a lightweight variant of U-Net during offline training. The input to the network consists of the sampled signal measurements $\Psi$ and the environmental map $\mathcal{M}$, and the output is the predicted radio map is given by
\begin{equation}
\hat{P} = f_{\mathrm{UNet}}(\Psi, \mathcal{M}; \theta).
\end{equation}

To enhance the reliability of radio map prediction, we incorporate uncertainty modeling into the U-Net framework. We distinguish between two types of uncertainty \cite{BNN}:  
\begin{itemize}
  \item \textbf{Aleatoric uncertainty} captures inherent noise in the observations, such as measurement variability or environmental fluctuations.
  \item \textbf{Epistemic uncertainty} captures the model’s uncertainty due to limited training data or ambiguous input conditions.
\end{itemize}
 Specifically, we estimate aleatoric uncertainty by extending the U-Net by adding an auxiliary output head to predict a spatially-varying noise variance map $\hat{\sigma}^2$ (see Fig.~\ref{system}(c)). The model is trained using \textit{negative log-likelihood loss}, under the assumption that each ground-truth pixel value $P_i$ is drawn from a Gaussian distribution with mean $\hat{P}_i$ and variance $\hat{\sigma}_i^2$ predicted by the network
\[
p\bigl(P_i \mid \hat{P}_i, \hat{\sigma}_i^{2}\bigr)
= \frac{1}{\sqrt{2\pi \hat{\sigma}_i^{2}}}\,
  \exp\!\Bigl[-\frac{(P_i-\hat{P}_i)^{2}}{2\,\hat{\sigma}_i^{2}}\Bigr],
\]
where \(p\) denotes a Gaussian density. The corresponding negative log-likelihood loss over all $N^2$ pixels is given by
\begin{equation}
\begin{aligned}
\mathcal{L}_{\mathrm{nll}}(\theta)
&= -\frac{1}{N^2} \sum_{i=1}^{N^2} \ln p(P_i \mid \hat{P}_i, \hat{\sigma}_i^2) \\
&= \frac{1}{N^2} \sum_{i=1}^{N^2} \left[
\frac{(P_i - \hat{P}_i)^2}{2\hat{\sigma}_i^2}
+ \frac{1}{2}\ln(2\pi \hat{\sigma}_i^2)
\right].
\end{aligned}
\label{eq:nll}
\end{equation}
This loss encourages the network to predict both accurate mean estimates $\hat{P}_i$ and meaningful uncertainty estimates $\hat{\sigma}_i^2$. A higher predicted variance reduces the penalty for residual error, allowing the model to express low confidence in ambiguous or under-sampled regions.

During inference, we estimate both aleatoric and epistemic uncertainties by combining heteroscedastic regression with MC Dropout. Specifically, we perform $K$ stochastic forward passes with dropout enabled and obtain $K$ predictions $\hat{P}_k = f_{\mathrm{UNet}}(\Psi, \mathcal{M}; \theta_k)$. We then compute the mean prediction $\bar{P} = \frac{1}{K}\sum_{k=1}^{K} \hat{P}_k$ as the final reconstructed radio map. The specific choice of $K$ will be detailed later.
The epistemic uncertainty is estimated as the variance across these stochastic predictions, while the aleatoric uncertainty is estimated by averaging the predicted variance maps: $\frac{1}{K}\sum_{k=1}^{K} \hat{\sigma}_k^2$. Finally, the total uncertainty map is computed by combining these estimates

\begin{equation}
U_K = \underbrace{\frac{1}{K} \sum_{k=1}^{K} (\hat{P}_k - \bar{P})^2}_{\text{Epistemic}} + \alpha \cdot \underbrace{\frac{1}{K} \sum_{k=1}^{K} \hat{\sigma}_k^2}_{\text{Aleatoric}},
\end{equation}
where $\alpha$ is a weighting factor that balances the contributions of these two uncertainties. The resulting uncertainty map $U$ provides pixel-wise estimates of information gain, which can be leveraged by the path planner to guide the agent toward the most informative regions for future measurements.

\subsection{Uncertainty-Aware Graph-Based Path Planning}
To enable adaptive exploration under a limited budget, we formulate the path-planning task as a sequential decision process over a weighted, sparsely connected graph constructed via PRM. Let $\mathcal{G} = (\mathcal{V}, \mathcal{E}, \omega)$ denote the roadmap, with $\mathcal{V}$ being a set of uniformly sampled, obstacle-free nodes across the environment. Each node is defined as $\psi_i = (v_i, \hat{P}_{v_i}, U_{v_i})$, where $v_i = (x_i, y_i)$ is the 2D coordinate, $\hat{P}_{v_i}$ is the predicted signal strength, and $U_{v_i}$ is the associated predictive uncertainty. The edge set $\mathcal{E}$ connects each node to its $k$ nearest neighbors $\mathcal{N}$ via collision-free paths, and $\omega$ assigns traversal costs proportional to the Euclidean distance between nodes, reflecting energy or time expenditure.
\setstretch{0.82}
Modeling the environment as a graph offers two key advantages:
\begin{itemize}
\item 
\textbf {Reflects Real-world Navigation:} It mirrors the waypoint-based navigation commonly used in real UAV/UGV systems, making it applicable to practical scenarios.
\item \textbf{Computationally Tractable Structure:} The graph provides a discrete structure that is suitable for sequential decision-making, allowing for efficient computation.
\end{itemize}
\setstretch{0.93}
 The agent begins at a designated start node $\psi_s$ and, at each step, selects its next move from the local neighborhood $\mathcal{N}(\psi_i)$ based on both the uncertainty values and traversal costs. This formulation enables efficient planning that balances exploration with resource constraints in a principled manner.

\begin{figure}[h]
    \centering
    \includegraphics[width=0.8\linewidth]{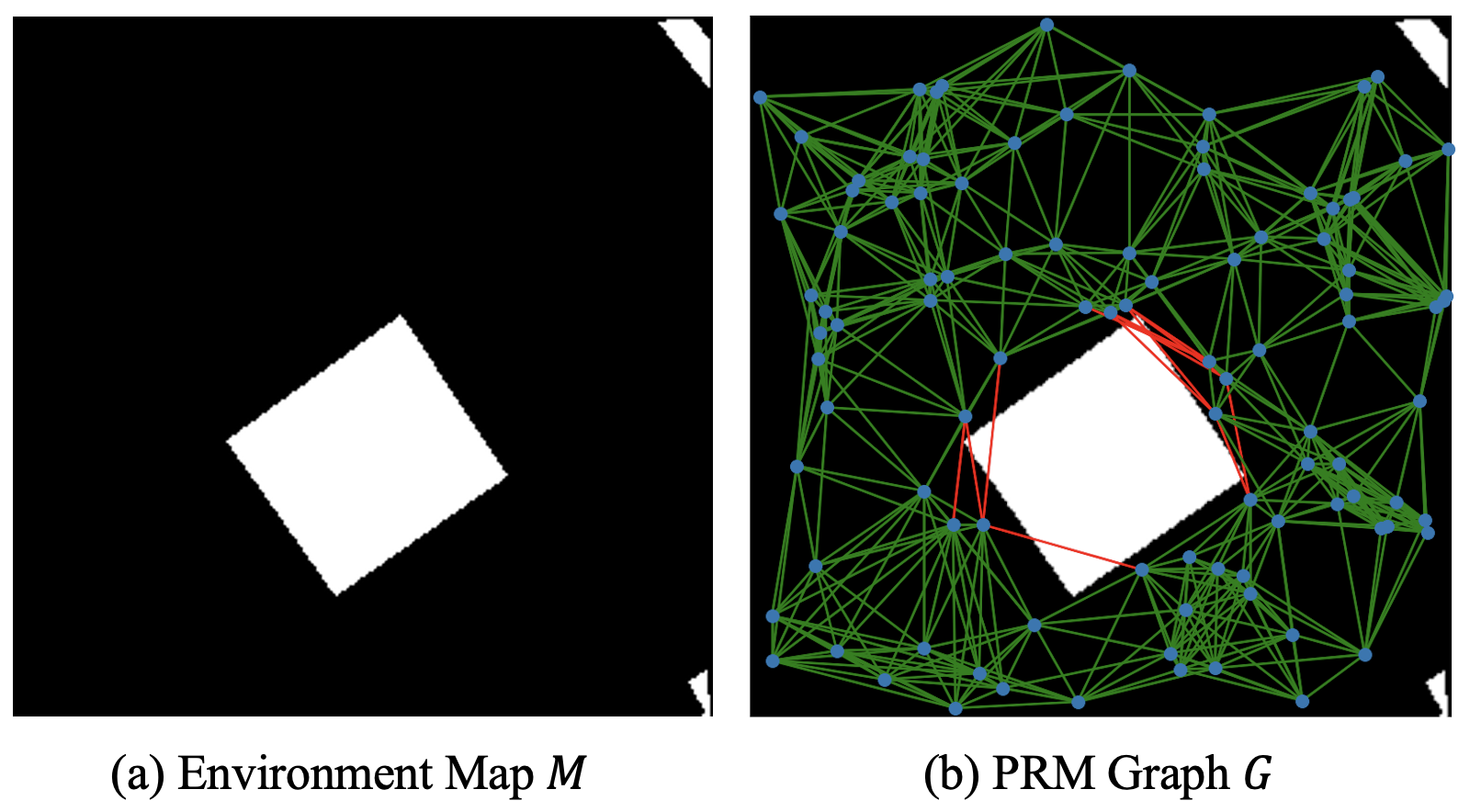}
    \caption{Construction of $\mathcal{G}$. \textbf{Left}: Binary obstacle map $\mathcal{M}$ where white areas indicate obstacles. \textbf{Right}: Generated PRM graph where nodes are sampled in free space, green edges represent collision-free connections, and red edges denote invalid connections intersecting obstacles.}
    \label{fig:graph}
\end{figure}

We construct $\mathcal{G}$ using the environment map $\mathcal{M}$, which encodes building obstacles (as Fig. \ref{fig:graph}). Define \( \mathcal{M}_{\text{free}} \subset \mathcal{M} \) as the set of obstacle-free regions in the environment map, where a straight line between two nodes is considered traversable if it lies entirely within \( \mathcal{M}_{\text{free}} \).
\newcommand{\seg}[2]{\operatorname{seg}\!\bigl(#1,#2\bigr)}  
For any edge $e_{ij}\!=\!(\psi_i,\psi_j)\in\mathcal{E}$, we define
\begin{equation}
w(e_{ij}) \;=\;
\begin{cases}
\dist{v_i}{v_j}, & \text{if } \seg{v_i}{v_j}\subseteq\mathcal{M}_{\mathrm{free}},\\[4pt]
B_{\max},        & \text{otherwise}.
\end{cases}
\label{eq:edge_weight}
\end{equation}
where $B_{\max}$ is a large constant penalizing occluded paths. This formulation ensures that the agent prefers paths through obstacle-free regions while heavily penalizing any path that intersects with obstacles, thereby promoting efficient and safe navigation within the environment.

The agent starts at node $\psi_s$, ends at node $\psi_g$ and interacts with the environment by selecting neighboring nodes to move to. A full trajectory is a sequence of nodes $\ (\psi_s, \psi_1, \dots, \psi_g)$, with movements executed along straight edges in $\mathcal{E}$. When transitioning from $\psi_m$ to $\psi_{m+1}$, the agent samples measurements along the entire line segment, collecting observations at all intermediate locations.

We model the path-planning sub-problem as a Markov Decision Process (MDP). At each decision step $m$, the components are as follows:
\setstretch{0.82}
\begin{itemize}
    \item \textbf{State $s_m$}: The state is defined as the agent’s current node $\psi_m$, the remaining budget $B_m$, and the roadmap graph $\mathcal{G} = (\mathcal{V}, \mathcal{E}, \omega)$. Each node $\psi \in \mathcal{V}$ includes its position, predicted signal strength, and uncertainty. 

    \item \textbf{Action $a_m$}: The action is defined as selecting the next node $\psi_{m+1} \in \mathcal{N}(\psi_m)$, where $\mathcal{N}(\psi_k)$ denotes the neighbors of $\psi_m$ in the graph. This corresponds to traversing an edge and acquiring new measurements along the associated path segment.

    \item \textbf{Reward $r_m$}: The reward is designed to encourage informative exploration by reducing predictive uncertainty across the entire graph. Let the total predictive uncertainty at step $t$ be defined as
\begin{equation}
\Sigma^{(m)} = \sum_{\psi \in \mathcal{V}} U_{v_i}^{(m)},
\end{equation}
where $U_{v_i}^{(m)}$ denotes the model-predicted uncertainty at node location $v_i$ at step $m$. The reward at each step is given by
\begin{equation}
r_m =
\begin{cases}
\frac{\Sigma^{(m-1)} - \Sigma^{(m)}}{\Sigma^{(m-1)}}, & \text{if } v_m \neq g, \cr
-\beta \cdot \Sigma^{(m)} & \text{if } v_m = g
\end{cases}
\end{equation}
The first case rewards the agent for reducing uncertainty from step $m-1$ to $m$,  as long as the current node location $v_m$ is not the goal node location $g$.
The second case applies a penalty proportional to the uncertainty $\Sigma^{(m)}$ if the agent reaches the goal node location $g$, with $\beta$ being a penalty factor.
This setup aims to strategically guide the agent's exploration by balancing the reduction of uncertainty with resource constraints, ultimately enhancing the efficiency and effectiveness of the exploration process.
\end{itemize}
\setstretch{0.87}
\subsection{Proximal Policy Optimization with Action Masking}
We train an attention-based policy $\pi_\theta(a_m\!\mid\!s_m)$ via PPO \cite{PPO} ,as illustrated in Fig.~\ref{system}(b). The attention mechanism is leveraged to capture graph-structured context and prioritize nodes with high uncertainty and strategic connectivity.

At each decision step $m$, we construct a binary action mask $\text{mask}_m$ as a vector with a dimension equal to the number of neighbors $|\mathcal{N}(\psi_m)|$ of the current node $\psi_m$. This mask is defined as
\begin{equation}
  \text{mask}_m \in \{0,1\}^{|\mathcal{N}(\psi_m)|}  
\end{equation}
The construction of this mask accounts for the residual budget $B_m$, collision checks, and a return-to-goal constraint, prunes infeasible neighbors before the policy is sampled. 
Neighbors deemed infeasible receive mask value $0$; feasible ones receive $1$. 
This effectively prunes infeasible neighbors before the policy is sampled.

The masked policy is then defined as
\begin{equation*}
    \pi_\theta^{\text{masked}}(a_m \mid s_m)=
\operatorname{Softmax}\!\bigl(\log\pi_\theta(a_k\mid s_m)+\log\text{mask}_m\bigr),
\end{equation*}
where $\log\text{mask}_m=-\infty$ for invalid actions and $0$ otherwise, guaranteeing zero probability for unsafe moves during training and inference.

Let $\theta$ and $\theta_{\text{old}}$ denote the current and previous policy parameters, respectively. Define the likelihood ratio as
$
r_m(\theta)=
{\pi_\theta^{\text{masked}}(a_m\mid s_m)}/
     {\pi_{\theta_{\text{old}}}^{\text{masked}}(a_m\mid s_m)}
$
. 
To stabilize training, we compute the advantage denoted $\hat{A}_m$ using Generalized Advantage Estimation (GAE). GAE leverages a learned state-value function $V(s)$, provided by a critic network trained jointly with the policy.

The PPO objective maximizes a clipped surrogate function given by
$\mathcal{L}(\theta) = \mathbb{E}_m \big[ \min \big( r_m(\theta) \hat{A}_m,\ \mathrm{clip}(r_m(\theta), 1 - \epsilon, 1 + \epsilon)\ \hat{A}_m \big) \big]$.
${\mathbb{E}}_m$ denotes the empirical average over a batch of collected timesteps. This objective is typically combined into a composite loss function that includes terms for the critic and exploration:
\begin{equation}
\mathcal{L}(\theta) = \hat{\mathbb{E}}_m \Big[ \mathcal{L}^{\text{CLIP}}(\theta) - c_1 \mathcal{L}^{\text{VF}}(\theta) + c_2 S[\pi_\theta](s_m) \Big].
\end{equation}
$\mathcal{L}^{\text{VF}}(\theta)$ represents a squared-error loss used to train the critic's value function, while $S[\pi_\theta](s_m)$ is an entropy bonus on the policy's output to encourage exploration. The terms are balanced by coefficients $c_1$ and $c_2$.

\begin{table}[t]
\centering
\small
\caption{Training hyperparameters for URAM.}
\begin{tabularx}{\linewidth}{@{} >{\centering\arraybackslash}m{1.2cm} 
                             >{\centering\arraybackslash}X 
                             >{\centering\arraybackslash}X @{}}
\toprule
\textbf{Component} & \textbf{Parameter} & \textbf{Value} \\
\midrule
\multirow{5}{*}{BNN} 
  & Batch size         & $64$ \\
  & Optimizer          & AdamW \\
  & Learning rate      & $1\times10^{-4}$ \\
  & Epochs             & $500$ \\
  & Dropout prob.      & $0.1$ \\
\midrule
\multirow{4}{*}{RL} 
  & Batch size         & $2048$ \\
  & Optimizer          & Adam \\
  & Learning rate      & $1\times10^{-4}$ \\
  & LR scheduler       & $step=32, \gamma=0.96$ \\
\bottomrule
\end{tabularx}
\label{tab:train_params}
\end{table}

\setstretch{0.80}
\subsection{Prediction and Planning Cycle}
\label{ppc}
The core of our methodology is an iterative prediction and planning cycle. At a given step $m$, the agent leverages a Bayesian U-Net, conditioned on the set of collected signal samples $\Psi^{(m)}$ and the physical environment map $\mathcal{M}$, to reconstruct the radio map $\hat{P}^{(m)}$. Concurrently, the network estimates a pixel-wise uncertainty map $U^{(m)}$ via MC dropout. These outputs, combined with the agent's current position, are passed to a self-attention-based path planner. The planner's objective is to identify a future trajectory that maximizes the expected reduction in global uncertainty while adhering to budget constraints. After executing the planned trajectory, the newly acquired samples are integrated into $\Psi^{(m)}$, forming the updated set $\Psi^{(m+1)}$, which initiates the next iteration of the cycle. This process repeats until the mission budget is depleted or the map's accuracy reaches the desired level.

\begin{figure}[h]
\centering
\includegraphics[width=\linewidth]{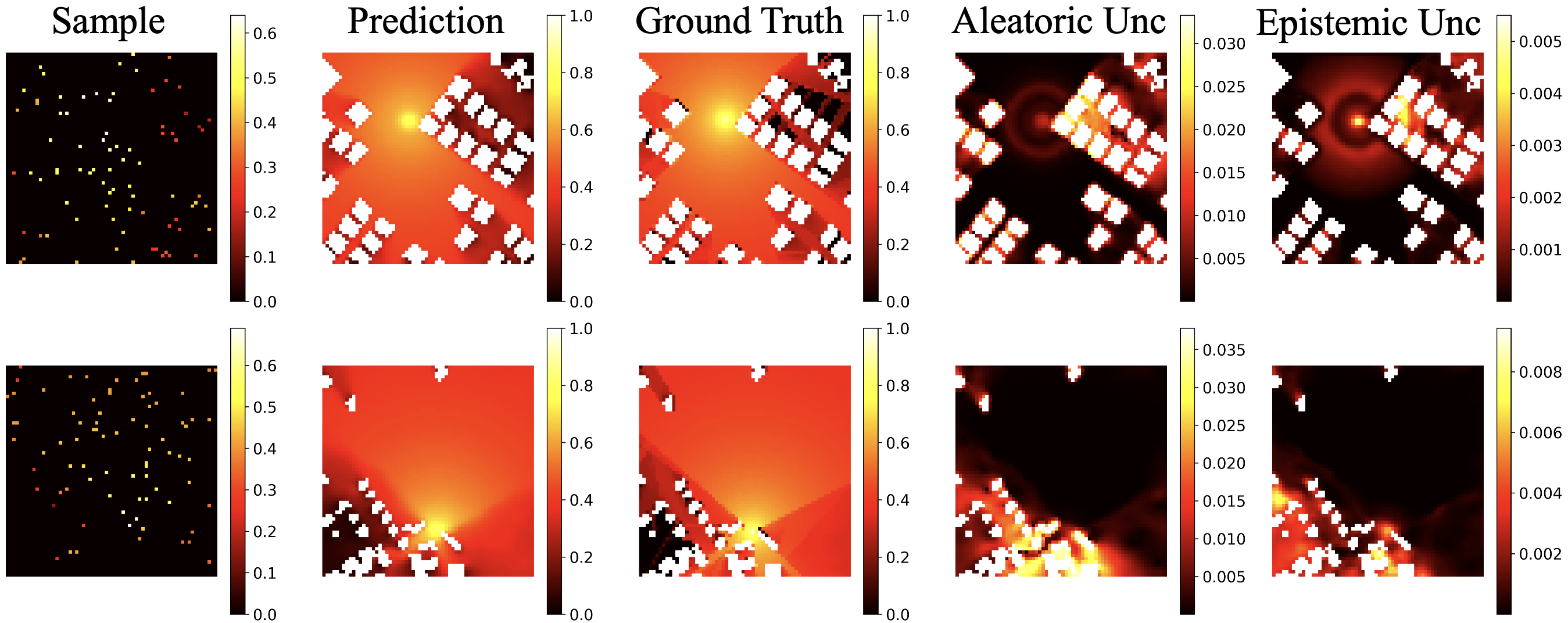} 
\caption{Visualization of the model output and uncertainty estimates on two test scenarios. The model produces higher uncertainty in complex regions.}
\label{fig:uncertainty_vis}
\end{figure}

The runtime of each URAM decision step is dominated by two main components: (i) the uncertainty-aware map prediction, which requires $K$ forward passes of the U-Net on an $N \times N$ map, leading to a complexity of $O(KN^2)$; and (ii) the path planner, whose self-attention mechanism over $|\mathcal{V}|$ graph nodes results in a complexity of $O(|\mathcal{V}|^2)$. The total complexity is therefore $O(KN^2 + |\mathcal{V}|^2)$. Through empirical evaluation, we determined that setting $K=10$ provides a robust and stable uncertainty map suitable for effective path planning, while avoiding the diminishing returns and higher computational burden of larger values. In contrast, reconstruction network like DRUE \cite{spectrumsurveying} operates at a faster $O(N^2)$ but it cannot quantify epistemic uncertainty. On the other hand, traditional methods like GPs provide exact variance but at an intractable cost of $O(N_s^2 + N_s N^2)$, where $N_s$ is the number of samples. This makes GPs impractical for real-time applications with large datasets ($N > 10^3$). URAM with a small constant $(K=10)$, thus provides a superior balance, delivering robust uncertainty estimates with computational costs that remain competitive with deterministic methods and significantly more scalable than GPs.

\section{Experiment}
\setstretch{0.78}
To validate the URAM framework, all experiments are conducted on the IRT2 subset of the RadioMapSeer dataset \cite{radiomapseer}, a second-order Intelligent Ray Tracing (IRT) corpus that emulates realistic multipath propagation in urban settings. IRT2 comprises 700 distinct $256\text{ m}\times256\text{ m}$ scenarios with heterogeneous building layouts and densities; each scenario is paired with 80 transmitter configurations, and the RSS is tabulated on a dense grid covering the entire area. From these 56000 scene–transmitter pairs we carve out 100 held-out environments—ten representative scenarios, each with ten transmitters—for policy training and benchmarking. This split guarantees diverse yet controlled test conditions for both radio-map reconstruction and uncertainty-aware path planning. The Root Mean Square Error (RMSE) is adopted as the error metric.

\subsection{Uncertainty-Aware Network}

To train the radio map prediction network, we downsample all radio maps in the RadioMapSeer dataset to a spatial resolution of $64 \times 64$. The entire dataset is split into training, validation, and test sets in a 7:2:1 ratio.

During training, each input consists of a sparse set of ground-truth samples, with a masking ratio randomly chosen between 95\% and 99\%. The target output is the full ground-truth radio map. The network is trained to reconstruct the complete radio map from sparse inputs while also estimating pixel-wise uncertainty.

\begin{table*}[!t]
\centering
\small
\caption{RMSE under different planning and reconstruction model combinations.}
\label{tab:baseline_matrix}
\begin{tabular}{cccc ccc ccc}
\toprule
\multirow{2}{*}{\textbf{Planning Method}} &
\multicolumn{3}{c}{\textbf{GP-DKL}} &
\multicolumn{3}{c}{\textbf{DRUE}} &
\multicolumn{3}{c}{\textbf{Bayesian U-Net (Proposed)}} \\
\cmidrule(lr){2-4} \cmidrule(lr){5-7} \cmidrule(lr){8-10}
& Best & Avg & Worst & Best & Avg & Worst & Best & Avg & Worst \\
\midrule
Mask-PPO (Proposed) & 0.401 & 0.441 & 0.532 & 0.036 & 0.041 & 0.048 & \textbf{0.026} & \textbf{0.030} & \textbf{0.037} \\
CMA-ES     & 0.412 & 0.452 & 0.545 & 0.033 & 0.046 & 0.053 & 0.032 & 0.036 & 0.044 \\
MCTS       & 0.420 & 0.448 & 0.539 & 0.031 & 0.049 & 0.056 & 0.031 & 0.035 & 0.048 \\
Random     & 0.382 & 0.464 & 0.580 & 0.042 & 0.063 & 0.069 & 0.026 & 0.047 & 0.057 \\

\bottomrule
\end{tabular}

\end{table*}

To estimate epistemic uncertainty, we adopt MC Dropout by enabling stochastic forward passes during inference. This approach approximates a Bayesian ensemble and provides a practical means of quantifying model uncertainty. The training configuration is summarized in Table~\ref{tab:train_params}.

We further visualize the model's predictions and associated uncertainty estimates in Fig.~\ref{fig:uncertainty_vis}, where both aleatoric and epistemic uncertainties are shown alongside the predicted radio maps. These results confirm that uncertainty is higher near unobserved or structurally complex regions, and that MC Dropout captures model uncertainty effectively.

\begin{figure}[t]
\centering
\includegraphics[height=5cm]{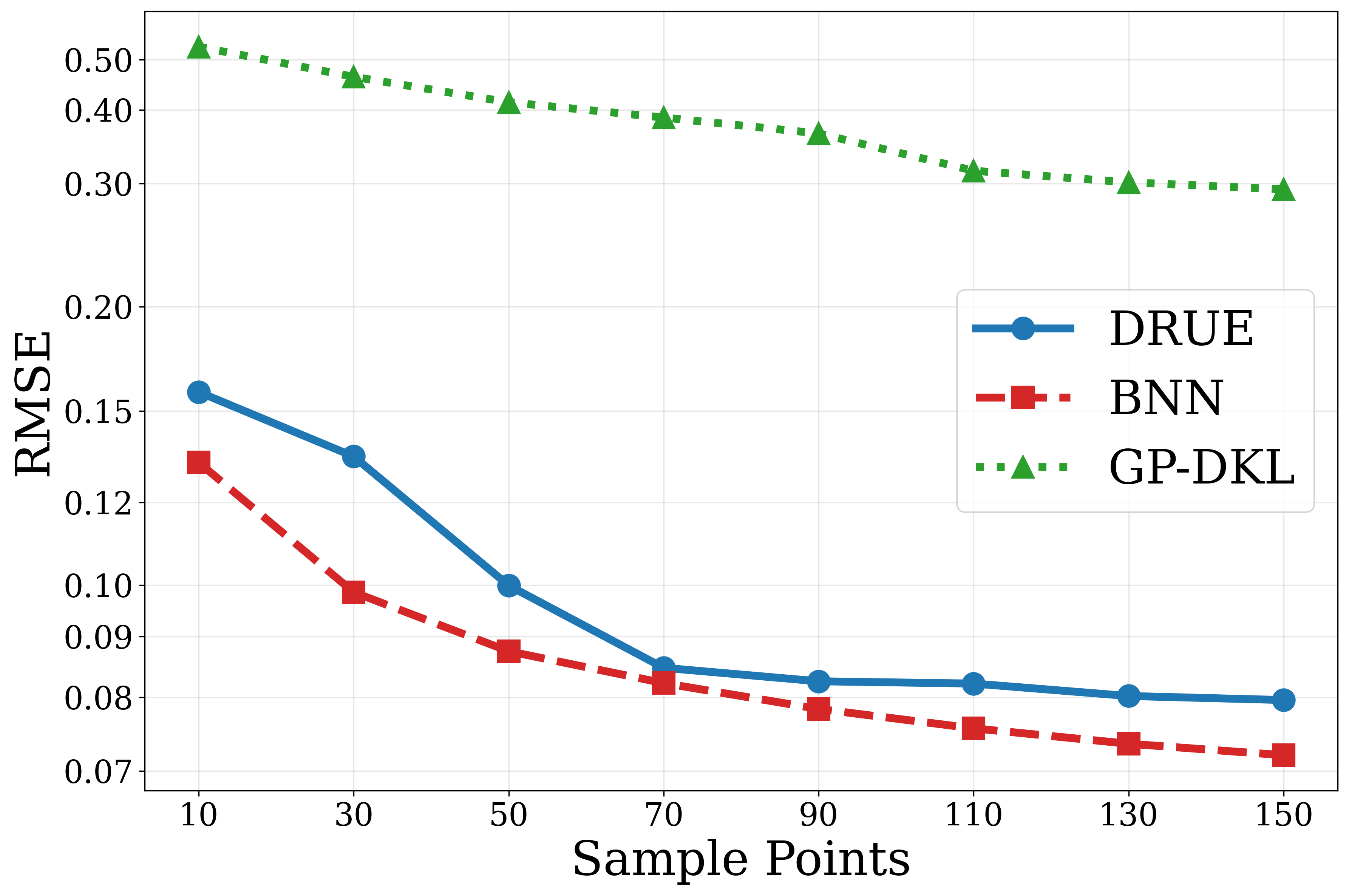}
\caption{Qualitative comparison of radio map reconstruction across Bayesian U-Net (ours), DRUE (autoencoder), and GP-DKL. }
\label{fig:comparison_vis}
\end{figure}

To assess reconstruction accuracy, we compare our Bayesian U-Net with two baselines: DRUE~\cite{spectrumsurveying}, which adopts an autoencoder architecture, and GP-DKL~\cite{dkl}, a Gaussian Process model with deep kernel learning. As shown in Fig.~\ref{fig:comparison_vis}, our U-Net-based model consistently yields the most accurate reconstructions, benefiting from its encoder–decoder design with skip connections. DRUE achieves moderate performance due to its simple autoencoder structure. GP-DKL performs worst in sparse regions, limited by scalability and kernel expressiveness.

We also compare the model params and time consuming of all methods, as summarized in Table~\ref{tab:inference_time}. These results empirically validate our complexity analysis. Bayesian U-Net with $K=10$ provides robust uncertainty estimation at a computational cost that is competitive with deterministic methods such as DRUE, and is more scalable than GP-based approaches. Although DRUE has the largest number of parameters, its inference is relatively faster due to its simple architecture and the fact that it runs only once during inference, whereas Bayesian U-Net requires multiple forward passes for uncertainty estimation. Overall, the Bayesian U-Net achieves a favorable trade-off between inference speed and reconstruction accuracy.

\begin{table}[t]
\small
\centering
\caption{Model Params and time comparison (measured on RTX 4090).}
\label{tab:inference_time}
\begin{tabular}{cccc}
\toprule
Method & Params (M) &  Time (ms) \\
\midrule
BNN ($K=10$) & 39.79 & 26.71 \\
DRUE & 118.28 & 22.73 \\
GP-DKL & 32.03 & 33.93 \\
\bottomrule
\end{tabular}
\end{table}

\subsection{RL Path Planning }

We train our Mask-PPO agents on the IRT2 subset of the RadioMapSeer dataset. A total of ten held-out scenarios, each with ten transmitter locations, result in 100 distinct planning environments for training. At the beginning of each episode, a PRM is constructed by randomly sampling 200 to 400 nodes. The trajectory budget is uniformly sampled between 80 and 150 waypoints. During rollout, the agent traverses the PRM and collects observations not only at the nodes but also at every intermediate point along each edge. The training parameters are summarized in Table~\ref{tab:train_params}.

To examine how the learned exploration policy cooperates with various reconstruction back-ends, we pair the trained Mask-PPO agent with three representative predictors—GP-DKL \cite{dkl}, DRUE \cite{spectrumsurveying}, and our Bayesian U-Net—and benchmark this combination against three alternative planning strategies: CMA-ES \cite{CMA}, Monte-Carlo Tree Search (MCTS) \cite{MTCS}, and random sampling. The resulting performance is expressed in terms of best-case, average, and worst-case RMSE, as presented in Table \ref{tab:baseline_matrix}.

As shown in Table~\ref{tab:baseline_matrix}, our BNN-based reconstruction model consistently achieves lower RMSE under all planning methods under budget $B=150$. In particular, the combination of BNN and RL yields the best overall performance, demonstrating both high sample efficiency and reconstruction fidelity. DRUE outperforms GP-DKL in most cases due to its stronger representation capacity. GP-DKL struggles in test scenarios due to kernel limitations and poor scalability.

\begin{figure}[t]
    \centering
    \includegraphics[width=\linewidth]{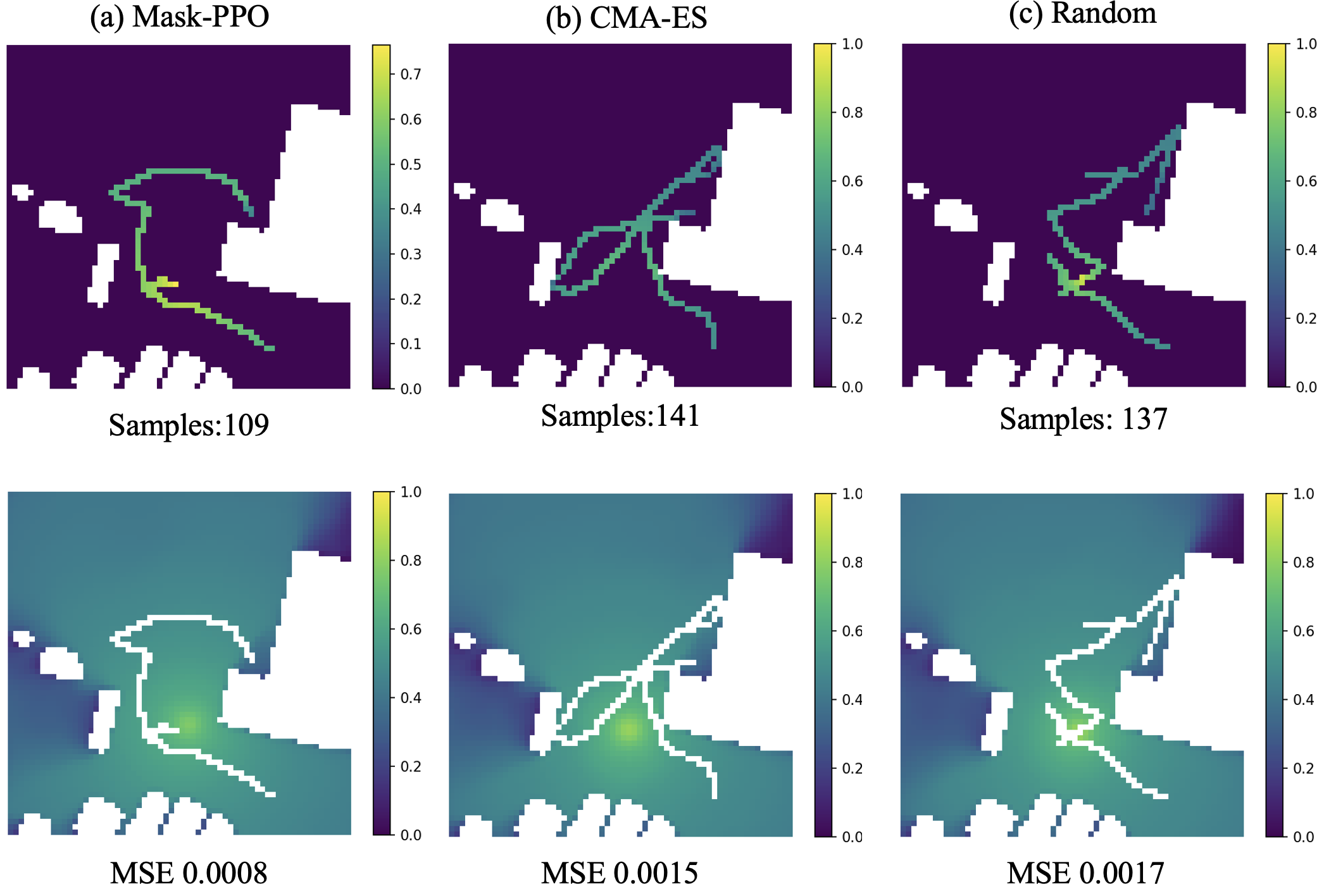}
    \caption{
    Comparison of the proposed RL method, CMA-ES and the random method.
    }
    \label{fig:rl_vs_greedy}
\end{figure}

We further provide a qualitative comparison in Fig.~\ref{fig:rl_vs_greedy}, visualizing one representative test case across three planning methods: Mask-PPO-based (proposed), CMA-ES, and random sampling. Each column shows the collected measurements (top) and the corresponding radio map reconstruction and residuals (bottom).

Despite using fewer samples (109 compared to 141 and 137), the proposed method achieves the most accurate reconstruction, yielding the lowest RMSE of 0.0282. Its trajectory effectively focuses on informative regions while avoiding redundant exploration. In contrast, the heuristic method collects more samples but yields higher error RMSE of 0.0387, and the random strategy performs the worst RMSE of 0.0412, exhibiting scattered observations and poor reconstruction in high-signal regions. This example highlights the efficiency and precision of our uncertainty-aware RL planner.

\section{Conclusions and Future Work}
\setstretch{0.8}
This paper has presented URAM, a novel framework that integrates bayesian uncertainty with deep reinforcement learning to enable autonomous, constraint-aware radio map construction. By training an agent to learn an intelligent, non-myopic exploration policy on a practical graph-based representation, our method creates higher-fidelity maps with greater budget efficiency than traditional heuristic baseline. While this work provides a robust foundation, future efforts will focus on extending the framework to more complex 3D and dynamic environments and deploying it on real-world aerial platforms.
\setstretch{0.7}
\bibliographystyle{IEEEtran}
\bibliography{Latex-2025/mybib}

\begin{thebibliography}{10}
\providecommand{\url}[1]{#1}
\csname url@samestyle\endcsname
\providecommand{\newblock}{\relax}
\providecommand{\bibinfo}[2]{#2}
\providecommand{\BIBentrySTDinterwordspacing}{\spaceskip=0pt\relax}
\providecommand{\BIBentryALTinterwordstretchfactor}{4}
\providecommand{\BIBentryALTinterwordspacing}{\spaceskip=\fontdimen2\font plus
\BIBentryALTinterwordstretchfactor\fontdimen3\font minus \fontdimen4\font\relax}
\providecommand{\BIBforeignlanguage}[2]{{%
\expandafter\ifx\csname l@#1\endcsname\relax
\typeout{** WARNING: IEEEtran.bst: No hyphenation pattern has been}%
\typeout{** loaded for the language `#1'. Using the pattern for}%
\typeout{** the default language instead.}%
\else
\language=\csname l@#1\endcsname
\fi
#2}}
\providecommand{\BIBdecl}{\relax}
\BIBdecl

\bibitem{ckm}
Y.~Zeng, J.~Chen, J.~Xu, D.~Wu, X.~Xu, S.~Jin, X.~Gao, D.~Gesbert, S.~Cui, and R.~Zhang, ``{A Tutorial on Environment-Aware Communications via Channel Knowledge Map for 6G},'' \emph{IEEE Communications Surveys \& Tutorials}, vol.~26, no.~3, pp. 1478--1519, 2024.

\bibitem{radiomap_survey}
D.~Romero and S.-J. Kim, ``{Radio Map Estimation: A Data-Driven Approach to Spectrum Cartography},'' \emph{IEEE Signal Processing Magazine}, vol.~39, no.~6, pp. 53--72, 2022.

\bibitem{raytracing}
\BIBentryALTinterwordspacing
J.~Hoydis, S.~Cammerer, F.~A. Aoudia, A.~Vem, N.~Binder, G.~Marcus, and A.~Keller, ``{Sionna: An Open-Source Library for Next-Generation Physical Layer Research},'' 2023. [Online]. Available: \url{https://arxiv.org/abs/2203.11854}
\BIBentrySTDinterwordspacing

\bibitem{kriging}
D.~Mao, W.~Shao, Z.~Qian, H.~Xue, X.~Lu, and H.~Wu, ``{Constructing Accurate Radio Environment Maps with Kriging Interpolation in Cognitive Radio Networks},'' in \emph{2018 Cross Strait Quad-Regional Radio Science and Wireless Technology Conference (CSQRWC)}, 2018, pp. 1--3.

\bibitem{matrixc}
H.~Sun and J.~Chen, ``{Propagation Map Reconstruction via Interpolation Assisted Matrix Completion},'' \emph{IEEE Transactions on Signal Processing}, vol.~70, pp. 6154--6169, 2022.

\bibitem{autoencoder}
Y.~Teganya and D.~Romero, ``{Deep Completion Autoencoders for Radio Map Estimation},'' \emph{IEEE Transactions on Wireless Communications}, vol.~21, no.~3, pp. 1710--1724, 2022.

\bibitem{nerf}
X.~Zhao, Z.~An, Q.~Pan, and L.~Yang, ``{NeRF2: Neural Radio-Frequency Radiance Fields},'' in \emph{Proceedings of the 29th Annual International Conference on Mobile Computing and Networking}, 2023, pp. 1--15.

\bibitem{payload}
V.~Semkin, S.~Kang, J.~Haarla, W.~Xia, I.~Huhtinen, G.~Geraci, A.~Lozano, G.~Loianno, M.~Mezzavilla, and S.~Rangan, ``{Lightweight UAV-based Measurement System for Air-to-Ground Channels at 28 GHz},'' in \emph{2021 IEEE 32nd Annual International Symposium on Personal, Indoor and Mobile Radio Communications (PIMRC)}, 2021, pp. 848--853.

\bibitem{activeGP}
K.~D. Polyzos, A.~Sadeghi, W.~Ye, S.~Sleder, K.~Houssou, J.~Calder, Z.-L. Zhang, and G.~B. Giannakis, ``{Bayesian Active Learning for Sample Efficient 5G Radio Map Reconstruction},'' \emph{IEEE Transactions on Wireless Communications}, vol.~23, no.~12, pp. 19\,382--19\,396, 2024.

\bibitem{ipp}
M.~Popovi{\'c}, T.~Vidal-Calleja, G.~Hitz, J.~J. Chung, I.~Sa, R.~Siegwart, and J.~Nieto, ``{An Informative Path Planning Framework for UAV-based Terrain Monitoring},'' \emph{Autonomous Robots}, vol.~44, no.~6, pp. 889--911, 2020.

\bibitem{spectrumsurveying}
R.~Shrestha, D.~Romero, and S.~P. Chepuri, ``{Spectrum Surveying: Active Radio Map Estimation with Autonomous UAVs},'' \emph{IEEE Transactions on Wireless Communications}, vol.~22, no.~1, pp. 627--641, 2023.

\bibitem{oreman}
N.~C. Matson and K.~Sundaresan, ``{Online Radio Environment Map Creation via UAV Vision for Aerial Networks},'' in \emph{2024 IEEE Conference on Computer Communications (INFOCOM)}, 2024, pp. 81--90.

\bibitem{cao2022catnipp}
Y.~Cao, Y.~Wang, A.~Vashisth, H.~Fan, and G.~Sartoretti, ``{Context-Aware Attention-based Network for Informative Path Planning},'' in \emph{Proceedings of The 6th Conference on Robot Learning}, 2023, pp. 1928--1937.

\bibitem{prm}
L.~Kavraki, P.~Svestka, J.-C. Latombe, and M.~Overmars, ``{Probabilistic Roadmaps for Path Planning in High-dimensional Configuration Spaces},'' \emph{IEEE Transactions on Robotics and Automation}, vol.~12, no.~4, pp. 566--580, 1996.

\bibitem{sip2net}
W.~Lu, Z.~Lu, J.~Yan, and S.~Gao, ``{SIP2Net: Situational-Aware Indoor Pathloss-Map Prediction Network for Radio Map Generation},'' in \emph{2025 IEEE International Conference on Acoustics, Speech and Signal Processing (ICASSP)}, 2025, pp. 1--2.

\bibitem{BNN}
A.~Kendall and Y.~Gal, ``{What Uncertainties Do We Need in Bayesian Deep Learning for Computer Vision?}'' in \emph{Proc. Advances in Neural Information Processing Systems}, vol.~30, 2017.

\bibitem{PPO}
\BIBentryALTinterwordspacing
J.~Schulman, F.~Wolski, P.~Dhariwal, A.~Radford, and O.~Klimov, ``{Proximal Policy Optimization Algorithms},'' \emph{CoRR}, vol. abs/1707.06347, 2017. [Online]. Available: \url{http://arxiv.org/abs/1707.06347}
\BIBentrySTDinterwordspacing

\bibitem{radiomapseer}
\BIBentryALTinterwordspacing
Çağkan Yapar, R.~Levie, G.~Kutyniok, and G.~Caire, ``{Dataset of Pathloss and ToA Radio Maps With Localization Application},'' 2024. [Online]. Available: \url{https://arxiv.org/abs/2212.11777}
\BIBentrySTDinterwordspacing

\bibitem{dkl}
A.~G. Wilson, Z.~Hu, R.~R. Salakhutdinov, and E.~P. Xing, ``{Stochastic Variational Deep Kernel Learning},'' in \emph{Proc. Advances in Neural Information Processing Systems}, vol.~29, 2016.

\bibitem{CMA}
\BIBentryALTinterwordspacing
N.~Hansen, ``{The CMA Evolution Strategy: A Tutorial},'' \emph{CoRR}, vol. abs/1604.00772, 2016. [Online]. Available: \url{http://arxiv.org/abs/1604.00772}
\BIBentrySTDinterwordspacing

\bibitem{MTCS}
\BIBentryALTinterwordspacing
M.~Swiechowski, K.~Godlewski, B.~Sawicki, and J.~Mandziuk, ``{Monte Carlo Tree Search: A Review of Recent Modifications and Applications},'' \emph{CoRR}, vol. abs/2103.04931, 2021. [Online]. Available: \url{https://arxiv.org/abs/2103.04931}
\BIBentrySTDinterwordspacing

\end{thebibliography}


\end{document}